%Paper: gr-qc/9310034
%From: brill@umdhep.umd.edu
%Date: Mon, 25 Oct 1993 17:48:02 EDT

\font\lbf=cmbx10 scaled\magstep2

\def\bs{\bigskip}
\def\ms{\medskip}

\def\ni{\noindent}
\def\cl{\centerline}

\def\title#1{\cl{\lbf #1}\ms}
\def\ctitle#1{\bs\cl{\bf #1}\par\nobreak\ms}
\def\stitle#1{\bs{\ni \bf #1}\par\nobreak\ms}

\def\ref#1#2#3#4{#1\ {\it#2\ }{\bf#3\ }#4\par}
\def\ns{\kern-.33333em}

\def\PR{Physical Review}

\magnification=\magstep1

\line{\hfill UMD 94-52}
\title{Comments on Initial Value Formulation}
\ctitle{Dieter R. Brill}
\cl{Department of Physics}
\cl{University of Maryland}
\cl{College Park MD 20742 U.S.A.}
\bs
\cl{Submitted to volume in {\it Einstein Studies} series}
\bs \bs
\ni {\bf Abstract.} This is the reply given at the conference
``Mach's Principle" at T\"ubingen in July 1993 to the paper by
Isenberg (1993a).
\ms\ni
\stitle{1. On Principles}\ni
Isenberg's (1993) proposal is remarkable not least because it
is intended to cover
not one or the other aspect of Machian ideas, but a complete formulation of
Mach's Principle. Isenberg gives cogent reasons why the Wheeler-Einstein-Mach
(W-E-M) program expresses the
important Machian demands, and it is hard to see how it could be
improved as a general program, particularly since Isenberg added
the nonextendibility requirement, giving a link between Mach's
principle and cosmic censorship.

Isenberg also considers Mach's principle in the larger context of
principles in physics. In this general context, Mach's principle
is somewhat unusual: It cannot easily be disproved, because we know
few if any effects that are unequivocally anti-Machian (for example,
Ozvath and Sch\"ucking 1962). By contrast, the most useful principles
in physics naturally have
a negative or interdictory aspect. For example, the uncertainty principle
forbids certain variables from being simultaneously well-defined, the
energy principle forbids perpetual motion, the equivalence principle
denies distinction between gravity and inertia, the atomic principle
excludes infinite divisibility, and so on. Such a formulation is not
only heuristically useful (for example, it saves us from useless speculation
about impossible situations) but it can also point the way toward
progress in the theory: a negative
principle implies a challenge, to find the mechanism or rationale behind
the prohibition, and can lead to a new theory in which the principle
is automatic, and no longer needs to be stated explicitly.

At first sight the  W-E-M principle
looks like business as usual  (we still do classical general
relativity in a way that current lingo might associate with STINO*),
and gives no direct motivation to change the theory.
The implication is different if we state it
as a negation: No spacetime can fail to satisfy the four W-E-M
requirements. But of course there {\it are}
solutions of the classical Einstein equations that are not W-E-Machian.
Hence the challenge is  to find the mechanism
that excludes the offending spacetimes. Thus the W-E-M principle also
points the way beyond classical general relativity to new and certainly as yet
unfinished business.

\stitle{2. On Inertia}\ni

Many, like Einstein, find something
fascinating about the idea that in inertia we feel the rest of the
universe at work, and look to Mach's principle for the real origin of inertia.
Does the W-E-M-Isenberg approach finish that business, of
formulating the principle?
Isenberg tells us that if we know the full spacetime metric near a point,
we know all there is to know about inertial frames at that point.
In Shimony's (1992) comparison, you enter
Mach's Store looking on the shelves for various useful
and fascinating gadgets, many of them somehow connected with inertia.
But in the W-E-M store you find only a general do-it-yourself kit from
which you might be able to build your own gadgets.
How much more effort is required to build the
gadgets we care about out of the W-E-M kit?
Let us consider some of the ``gadgets" that other authors in this
volume might hope to find in Mach's store.

Prof.~Pfister might care about the inertial frame dragging.
Suppose we
consider a point inside Pfister's shell. We know the metric there ---
it is flat. But this knowledge does not tell us all there is to know about
the dragging as usually understood
(Brill and Cohen 1966, Lindblom and Brill 1974). A true answer about inertia
and inertial frames must involve specific frames or coordinates.
The W-E-M principle, being a child of general relativity,
tends to be hostile to picking
out a particular frame --- the really significant information is
considered to be frame-independent. ``Frame not included"
is written on the packages in the W-E-M store; but is this not one of
the things we expect to get from Mach, not to put into it?

Prof.~Raine, whose own formulation of Mach's principle has been
questioned concerning the distinction between matter and gravitational
waves, might ask of the W-E-M principle whether there is
really a crucial difference between the following two situations:  an
otherwise closed W-E-M universe containing either a black hole
formed by collapse of matter, or an eternal Kruskal black hole,
with an asymptotically flat region on the ``other side" of the horizon.
The former would be called W-E-Machian, and the latter would not, because
its $\Sigma^3$ is not compact. But this distinction is not reasonable: since
the difference can be extremely small between the physical regions on ``this
side" of the horizon, and since one cannot look behind a horizon, the
Machian nature
of a spacetime would be something that could never be ascertained by
experiment. Perhaps the attribute
{\it Machian} should apply to {\it regions} in spacetimes, for which it does
not matter what happens behind horizons.

If we allow this extension of the W-E-M principle we can treat the
following situation, which is more amusing than
profound.  Suppose Prof.~Narlikar asked the question that has a definite
answer in his formulation: what is the smallest number of masses in
a W-E-Machian $S^3$ universe that is free from other content such as
gravitational waves? Suppose we take the absence of wave content to mean
that the free data can be chosen to be trivial, and the presence of
mass to mean that $n$ asymptotically
flat regions behind (apparent) horizons are allowed. Since asymptotically
flat regions are conformally equivalent to taking points out of the $S^3$,
an appropriate choice for Isenberg's first set ($\Sigma^3,\ \lambda,\
\sigma$) is
$({\rm I\!R}^3$ less $(n-1)$ points, flat, $0)$.
For $n=1$ the only regular solution for the Lichnerowicz conformal factor
$\phi$ is $\phi =$ constant, which is flat spacetime without horizon,
hence without Machian region. For $n=2$ the solution is
$\phi = 1 + {M \over {2r}}$, with $r$ = Euclidean distance in
${\rm I\!R}^3$ from the removed point. This is just the single-mass
Schwarzschild solution with one horizon, which does not bound a compact
Machian region. So one mass is not enough.
For $n=3$ we have $\phi = 1 + {M_1 \over {2r_1}} + {M_2 \over {2r_2}}$,
which is asymptotically flat in three regions, at $r_1 \rightarrow \infty$,
at $r_2 \rightarrow \infty$, and at $(r_1$ and $r_2) \rightarrow \infty$.
For small $M_1,\ M_2$ there are only two horizons, not bounding a
Machian region. But if $M_1$ and $M_2$ are chosen large enough
(compared to their Euclidean distance), there can be another apparent
horizon surrounding the two (Brill and Lindquist 1963). A Machian region
then exists between these three horizons. Thus three masses is the
answer by this extended W-E-M principle, not unreasonable because three
masses usually do define a frame. (Unfortunately in this particular
construction they do not, because the solution is rotationally symmetric
about an axis through the original $M_1$, $M_2$. In this sense the answer
is not better than Narlikar's two-mass minimum.)

\stitle{3. On Details}

Examples such as the above suggest that the W-E-M principle leaves
some room for further refinement. This appears particularly urgent
in connection with the distinction between the ``first"
and ``second" set of Cauchy data. The first set should contain variables
that can be freely
chosen; but in Isenberg's examples it consists of a $TT$ tensor $\sigma$ and
{\it transverse} fields $\beta$ and $\eta$. Because of such transversality
requirements these fields are really not
free, but themselves subject to constraints. Would it then not be
simpler to choose as the first set any constraint-satisfying
initial data, so that the second set is empty? If it is allowed to demand
transversality of the first set, then why not constraint-satisfaction?
Isenberg (1993b) suggests that the former condition is linear
and does not essentially restrict free choice,
whereas the latter
condition is non-linear and implements the Machian determination of
the inertial frames.  This interpretation itself would of course constitute a
(small) refinement  of the W-E-M principle, a refinement
motivated by a possible physical meaning of the
splitting into first and second sets.

Refining the physical meaning of the decomposition of data into the
first and second set seems a promising task. It could have
interesting physical significance if a {\it particular} decomposition
were demanded, not just the existence of {\it some} decomposition
(one of possibly many).  For example, in the Lichnerowicz-York decomposition,
the vector $W$ itself does not appear in the ``Machian" constraints,
only $LW$ occurs. Perhaps  this (or some other, even more Machian)
decomposition can give an appropriate, general definition of the frame dragging
by means of a vector like $W$ (which may be related to the shift vector,
a coordinate quantity of the type needed really to describe inertia).

It would seem unusual to find that a formulation, one of whose authors
is Wheeler, could benefit from greater emphasis on physical meaning,
but such are the conclusions to which we are led.

\bs\ni
{\bf Footnote}\ms\ni
*STINO, from {\it stinknormal}, name of new popular music phenomenon
in Germany, celebrating traditional melodies and folk songs. Perhaps
such labels can help us gain recognition in the lay public. (What
attention the no-hair theorems might have received if they were
identified with skinheads!)
\bs
\begingroup
\parindent=0pt\everypar={\global\hangindent=20pt\hangafter=1}\par
{\bf References}\ms
\ref{Brill, Dieter R. and Cohen, Jeffrey 1966 ``Rotating Masses and their
effects on inertial frames"}{\PR}{143}{1011-1015}
\ref{Brill, Dieter R. and Lindquist, Richard W. 1963 ``Interaction Energy
in Geometrostatics"}{\PR}{131}{471-476}
\ref{Isenberg, James 1993a ``On the Wheeler-Einstein-Mach Spacetimes"}
\ns\ns{this volume}
\ref{Isenberg, James 1993b}{private communication}\ns{T\"ubingen July 1993}
\ref{Lindblom, Lee and Brill, Dieter R. 1974 ``Inertial effects in the
gravitational collapse of a rotating shell"}{\PR}{D 10}{3151-3155}
\ref{Ozsv\'ath, Istv\'an and Sch\"ucking, Engelbert 1962 ``An Anti-Mach
Metric" in}
{Recent Developments in General Relativity}\ns{p.~339-350, Pergamon Press}
\ref{Shimony, Abner 1992}{private communication}\ns{APS Spring Meeting,
Washington, DC. Also see Brill, Dieter R. {\it Comments on Dragging
Effects}, this volume}
\endgroup

\end